\documentstyle[preprint,aps]{revtex}          
\begin{document}
\def\sla#1{\rlap\slash #1}
\title{Pairs in the light-front and covariance
\footnote{Nucl.Phys.{\bf A} (1998c): 
Procceedings of the XV Few Body Conference,
Groningen,1997.}}
\author{J.P.B.C. de Melo$^a$, J.H.O. Sales$^b$, 
T. Frederico$^b$ and P.U. Sauer$^c$}
\address{
$^a$ Instituto de F\'\i sica, Universidade de S\~ao Paulo, \\ 01498-970
S\~ao Paulo, S\~ao Paulo, Brazil. \\
$^b$ Dep.de F\'\i sica, Instituto Tecnol\'ogico de Aeron\'autica, 
Centro T\'ecnico Aeroespacial, \\
12.228-900 S\~ao Jos\'e dos Campos, S\~ao Paulo, Brazil.\\
$^c$Institute for Theoretical Physics, University Hannover \\ 
D-30167 Hannover, Germany}
\date{\today}
\maketitle
\begin{abstract}
The electromagnetic
current of bound systems in the light-front is constructed in
the Breit-Frame, in the limit of momentum transfer $q^+=(q^0+q^3)$ vanishing.
In this limit, the  pair creation term survives and it is
 responsible for the
covariance of the current. The pair creation term is computed for
the $j^+$ current of a spin one composite particle  in the Breit-frame.
The rotational symmetry of $j^+$ is violated if the pair term is 
not considered.
\end{abstract}
\pacs{12.39.Ki,14.40.Cs,13.40.Gp}

The electromagnetic current in the light-front can be constructed
from the covariant Feynman one-loop triangle diagram, 
which is integrated in the  $k^-(=k^0-k^3)$ component
of the internal loop momentum \cite{saw92,tob92,shakin,jaus}.
The  Cauchy integration just
includes the residue at the
pole of the forward propagating  spectator particle in
the photon absorption process for $q^+=0$.
It is understood that it is possible
to eliminate pairs created out or annihilating into the vacuum, which
leads to a description in terms of
of a two-particle light-front wave-function\cite{saw92,tob92,jaus}.
In general, this procedure keeps
the covariance under kinematical boost transformations, but 
the current loses its physical properties under general rotations and
parity transformations. As we will show, the pair terms 
should be evaluated for the full covariance of the current.

Long ago, the problems with the naive Cauchy integration of $k^-$
in momentum loops, which takes into account only the residue at the pole of
the forward propagating particle, had been addressed in Ref. \cite{yan}.
However, in that work, it was not discussed the electromagnetic current.
The purpose of our work is to derive  
the pair creation terms. They are responsible for the covariance of
the photon absorption process in the light-front.
We show how the pair terms maintain the rotational symmetry  of 
the  "good"-component of the electromagnetic current ($j^+$) \cite{dash}
in the Breit-frame for a spin one two-fermion composite system. In this
case the angular condition is valid.
The angular condition has been discussed in the context
of the Hamiltonian light-front dynamics \cite{frank,inna84,to93,keis94}.

We begin our discussion
with an illustrative example, where a pair term appears. 
We show that, the $j^-(=j^0-j^3)$ component
of the current of a two-boson bound state has contribution from
the pair term in the Breit-frame with momentum transfer such 
that $q^+=q^-=0$. This completes a previous work on 
the subject \cite{saw92}, where
$j^+$ and $ j_\perp$ have been calculated from the triangle diagram. 
For these components only the contribution from the residue 
corresponding to the forward propagating
boson survives in the Cauchy integration
over $k^-$ in the momentum loop. The calculation have been 
 done in the Breit-frame, where the momentum transfer $q^\mu$
is along the transverse direction $x$; $q^\mu=(0,q_x,0,0)$. No pair terms
survive for $j^+$ and $ j_\perp$, which is not valid for $j^-$.
The current $j^-$ is known to be the "bad"-component of the current
\cite{dash}, thus
in principle it receives contribution from the pair term. It is not apparent
how such term appears for $q^+=0$, and to obtain it, a careful
analysis of the limit $q^+\rightarrow 0$ is required.

The $j^-$-component of the electromagnetic current is given by:
\begin{eqnarray}
j^{-}=\Gamma^2 \int\frac{d^4k}{(2\pi)^4} \frac{ \   \ (2 k - P' - P)^{-}}
{(k^2 - m^2+\imath \epsilon)((k-P)^2 - m^2+\imath\epsilon) 
((k-P^{'})^2 - m^2+\imath \epsilon)} \ ,
\label{jboson}
\end{eqnarray}
where $P$ and $P'$ are the initial and final quadri-momenta of
the composite boson; $\Gamma$ is the vertex of the composite
state and $m$ is the boson mass.

The integral, Eq. (\ref{jboson}),  in the
light-front coordinates, $ k^{+} $ and $ k^{-}$,
is written as
\begin{eqnarray}
j^{-}&=&\Gamma^2\int\frac{d^{2} k_{\perp} d k^{+}d k^{-}}{2(2 \pi)^4} 
\frac{ (2 k - P' - P)^{-}}{k^+(P^+-k^+) (P^{' +}-k^+)(k^{-}-\frac{f_1-
\imath\epsilon}{k^+})} \\ \nonumber
& & \frac{}{(P^{-}-k^{-}-\frac{f_2-\imath\epsilon}{P^+-k^+}) 
(P^{'-}-k^{-}-\frac{f_{3}-\imath\epsilon}{P^{'+}-k^+})} \ ,
\label{jnpbos}
\end{eqnarray}
where
$ f_1 =  k^{2}_\perp+m^2  \ \ , \ \
f_2  =  (P-k)^{2}_\perp+m^2  \ \ , \ \
f_3  = (P^{'}-k)^{2}_\perp+m^2  \ . $
The Cauchy integration over $k^-$ in Eq.(\ref{jnpbos}) has two 
nonzero contributions to the residue
calculation, one coming from  the interval
$ 0 < k^{+} < P^{+}$ (I) and
another from $P^+ <k^+<P^{'+}$ (II). We use $q^+$ nonzero
with $P^{'+} = P^{+} + q^+$, and let then $q^+$ goes to zero, such that
the Breit-frame is recovered. Surprising, the contribution
 from (II) is nonzero in the above  limit.

The Cauchy integration in $k^-$ for $0<k^+<P^+$ (I), 
receives  contribution from the residue at the 
forward propagating on-mass-shell
pole of the spectator particle in the photon absorption process.
This pole is placed
in the lower semi-plane of complex-$k^-$ at
$k^-=(f_1-\imath \epsilon)/k^+$. It
is the only pole which contributes to the Cauchy integration
 in $j^+$ and $j_\perp$ 
in the limit of vanishing $q^+$ \cite{saw92}.
The Cauchy integration of $j^-$ in region (I) results
\begin{eqnarray}
j^{-I}&=& -\imath \Gamma^2 \int
\frac{d^{2} k_{\perp} d k^{+}}{2(2 \pi)^3} 
\frac{\left[ 2f_1- k^+(P^{'-}+P^{-})\right]
\theta(P^+-k^+)\theta(k^+)}{k^{+2}(P^+-k^+) (P^{' +}-k^+)
(P^{-}-\frac{f_1}{k^+}-\frac{f_2}
{P^+-k^+})(P^{'-}-\frac{f_1}{k^+}-\frac{f_3}
{P^{'+}-k^+})} \ . 
\label{ji}
\end{eqnarray}

The second contribution to the Cauchy integration of $k^-$ in Eq.(\ref{jnpbos})
comes from the region (II) where $P^+ < k^+ < P^{'+}$. The integration 
can be performed in the upper half of the complex-$k^- $ plane. Only 
the residue at the
pole $ k^- =P^{'-}-(f_3-\imath\epsilon)/(P^{'+}-k^+)$ contributes to
the Cauchy integration, 
\begin{eqnarray}
j^{-II}&=&-\imath \Gamma^2 
\int \frac{d^2k_{\perp} d k^{+}}{2(2 \pi)^3} 
\frac{\left[ -2\frac{f_3}{P^{'+}-k^+}+ P^{'-}- P^- \right]
\theta(P^{'+}-k^+)\theta(k^+-P^+)}
{k^+(P^+-k^+) (P^{'+}-k^+)
(P^{'-}-\frac{f_3}{P^{'+}-k^+}-\frac{f_1}{k^+}) } \\ \nonumber
& & \frac{}{(\frac{f_3}{P^{'+}-k^+}-\frac{f_2}{P^+-k^+}+P^- - P^{'-})}  
\  .
\label{jii}
\end{eqnarray}
The physical process represented by
Eq.(\ref{jii}) is the pair creation by the
photon. The denominator 
$\left[ \frac{f_3}{P^{'+}-k^+}-\frac{f_2}{P^+-k^+}+P^- - P^{'-}\right]^{-1}$
corresponds to the forward propagator of the virtual three-particle
system composed by the initial bound state, the antiparticle and
the particle produced by the incoming photon.
The denominator 
$\left[ P^{'-}-\frac{f_3}{P^{'+}-k^+}-\frac{f_1}{k^+}\right] ^{-1}$
is the forward propagator of the virtual two particle system,
 which composes the bound light-front wave-function in the final state.

Before performing the limit of $q^+\rightarrow 0_+$, we make the following
variable transformation, $x=(k^+-P^{+})/q^+$.  Eq. (\ref{jii}) becomes
\begin{eqnarray}
j^{-II}&=&\imath \Gamma^2 \int \frac{d^2k_{\perp} dx}{2(2 \pi)^3} 
\frac{\left[ -2 f_3+ q^+(1-x)(P^{'-}- P^-)\right]
\theta(1-x)\theta(x)}
{(P^++q^+x)x (1-x)^2(-\frac{f_3}{1-x}-q^+(\frac{f_1}
{P^++q^+x}-P^{'-}))} \\ \nonumber
& & \frac{}{(\frac{f_3}{1-x}+\frac{f_2}{x}+q^+(P^- - P^{'-}))}   \  .
\label{jiix}
\end{eqnarray}

In Eq.(\ref{jiix}) the limit $q^+\rightarrow 0_+$ can be done,
resulting
\begin{eqnarray}
j^{-II}&=& \imath 
\frac{\Gamma^2}{P^+}\int \frac{d^2k_{\perp} dx}{(2 \pi)^3}
\frac{\theta(1-x)\theta(x)}{x f_3+(1-x)f_2} =\imath\frac{\Gamma^2}{P^{+}}
\int  \frac{d^2 k_{\perp}}{(2 \pi)^3} 
\frac{\ln(f_3)-\ln(f_2)}{f_3-f_2 }   \  .
\label{jiixx}
\end{eqnarray}
The sum of $j^{-I} $ and $j^{-II}$ is equal to the covariant expression
Eq.(\ref{jboson}), $j^{-}=j^{-I} + j^{-II} $.

The pair term assures the correct
parity transformation properties of the current
in the light-front. The parity operator
for the transformation of $z$ into $-z$ is of non-kinematical
nature, consequently one expects that parity properties coming from this 
operation are destroyed, if pair terms are neglected. Below, we 
 illustrate this point. Consider the integral
\begin{eqnarray}
Z =\int \frac{d k^{+}d k^{-}}{2(2\pi)^2}
\frac{k^+ - k^-}{(k^2 - m^2+\imath \epsilon)((k-P)^2 - m^2+\imath\epsilon) 
((k-P^{'})^2 - m^2+\imath \epsilon)}
 \ ,
\label{z}
\end{eqnarray}
that vanishes in the Breit-frame where $q^+=q^-=0$. The argument 
of the integral is
odd under the transformation of $k^3\rightarrow -k^3$ ($k^3=(k^+-k^-)/2$).
However, if the Cauchy integration is performed in $k^-$
taking into account only the residue at 
the pole $k^-=(f_1-\imath \epsilon)/k^+$
for $0 < k^+<P^+$, it results nonzero
\begin{eqnarray}
Z = & &  \imath\int^{P^+}_0\frac{d k^{+}}{4\pi}
\frac{ k^+ - \frac{f_1}{k^+}}{k^+(P^+-k^+) (P^{ +}-k^+)
(P^{-}-\frac{f_1}{k^+}-\frac{f_2}{P^+-k^+})
(P^{-}-\frac{f_1}{k^+}-\frac{f_3}{P^{+}-k^+})} \nonumber \\
= & & \frac{\imath}{2\pi} \frac{\ln(f_2)-\ln(f_3)}{ P^+ (f_3-f_2)} \ .
\label{z+}
\end{eqnarray}
This example shows that it is dangerous to commute  the limit
$q^+\rightarrow 0_+$ with the Cauchy integration in $k^-$.
However, the last term in Eq.(\ref{z+}) is 
exactly  the opposite value of the residue that comes from region (II)
$( P^+ <k^+<P^{'+})$ in the limit of $P^{'+}\rightarrow P^+$. This contribution
is obtained from the same arguments used in deriving Eq.(\ref{jiixx}). 
By summing both residues  the above integration exactly cancels out,
as it should be  from  parity
considerations. The integral which has $k^-$ in the numerator can be
classified as "bad", since it has contribution from the pair creation
process.

The Eq.(\ref{jiixx}) can be generalized to include any number 
of denominators. This will be 
important for the study of the $j^+$-component of the
electromagnetic current for the composed vector
particle. The integral presented below 
corresponds to  the general form of the pair terms
\begin{eqnarray}
& & \lim_{q^+\rightarrow 0}\imath \int^{P^{'+}}_{P} 
\frac{d k^{+}(k^{+})^{n-1}}{2(2 \pi)} 
\frac{2 f_3}
{(P^+-k^+)^{N-2} (P^{'+}-k^+)^2(\frac{f_3}{P^{'+}-k^+}-\frac{f_1}{k^+})
\prod_{j\geq 2,j\neq 3}^N(\frac{f_3}{P^{'+}-k^+}
-\frac{f_j}{P^+-k^+})} \nonumber \\
&=&-\frac{ \imath}{2\pi} \sum_{i=2}^{N}\frac{ \ln(f_{i})}
{\prod_{j=2,i\neq j}^{N}(-f_i + f_j)} (P^+)^{n-1} \ ;
\label{crt}
\end{eqnarray}
in particular for $N=3 $ and $ n=0 $, Eq.(\ref{crt}) 
reduces to the integrand of Eq.(\ref{jiixx}).

Now we discuss  the
"good" component of the current of the composite
vector particle. For our purpose
of presenting the main points on how the pair terms maintain the
rotational properties of
 $j^+$  in the light-front, we have chosen 
a ${ \overline \psi }\epsilon^{\mu}_i \gamma_{\mu}\psi$ 
coupling. The four-vector
$\epsilon_i^{\mu}$ is the polarization of the vector particle, in the
$i(=x,y,z)$ direction. The covariant form of $j^+$ is given by
\begin{eqnarray}
 j^+_{ji}&=&\imath  \int\frac{d^4k}{(2\pi)^4}
\Lambda(k,P')\Lambda(k,P) 
\frac{Tr[\sla{\epsilon_j'}
(\sla{k}-\sla{P'} +m) \gamma^{+} 
(\sla{k}-\sla{P}+m) \sla{\epsilon_i}
(\sla{k}+m)]}
{((k-P)^2 - m^2+\imath\epsilon) 
(k^2 - m^2+\imath \epsilon)
((k-P')^2 - m^2+\imath \epsilon)}
 \ , \label{jvec}
\end{eqnarray}
where $j^+_{ji}$ is written in the Cartesian instant-form 
spin basis, and $\epsilon^{'\alpha}_j$ is the
 final polarization four-vector
 and $\epsilon^\beta_i$ is the initial
four-vector polarization.
The vector particle four-momentum  in the
Breit-frame are 
$P^\mu=(P^0,-q_x/2,0,0)$ for the initial state, 
and ${P^{'\mu}}=(P^0,q_x/2,0,0)$
for the final state;  $P^0=m_v \sqrt{1+\eta} $, where
$\eta\ = \ -q^2/4 m_v^2 $.

The regularization function,
$
\Lambda(k,P) =C\left[(k-P)^2 - m^2_R+\imath\epsilon\right]^{-1}
$ ,
was chosen to turn Eq.(\ref{jvec}) finite.
The special form of the regulator, allows to
identify a null-plane wave-function similar to the one
proposed for the pion in Ref.\cite{shakin}. The 
factor $C$ is fixed by the charge normalization.

The trace in the numerator for the different matrix elements
of the current, are written for the possible polarization states.
The "good" ones, denoted by $Tr^g_{ji}$, correspond to the
traces that have dependence only on $k^+$, $k_\perp$ and 
$k^-(P^+-k^+)^m$ with $m= 1,2,3...$. The last ones
do not contribute in the interval of $P^+< k^+<P^{'+}$
for $q^+\rightarrow 0_+$, because
the difference $(P^+-k^+)^m$ is of the order of $(q^+)^m$ which
vanishes  in this limit. For the "good" terms,
 only the residue at the spectator 
particle on-mass-shell pole contributes to the Cauchy 
integration over $k^-$ in the limit of $q^+\rightarrow 0_+$.
This pole in Eq.(\ref{jvec})
is placed on the lower
half of the  complex $k^-$ plane at $ k^-=(k^2_\perp+m^2-\imath
\epsilon)/k^+$ with $0< k^+ <P^+$.

The traces in the numerator of Eq.(\ref{jvec}), can be written in
the following form
\begin{eqnarray}
Tr_{xx} &=& Tr^g_{xx}- k^-\left( k_\perp^2 + m^2   - \frac{q_x^2}{4} \right)
\frac{q_x^2}{m^2_v} \ ; \
Tr_{zx}= Tr^{g}_{zx}- 2 k^-\left( k_\perp^2 + m^2-\frac{q_x^2}{4} \right)
\frac{q_x}{m_v} \ ; \
\nonumber
\\
Tr_{yy} &=& Tr^{g}_{yy} \ \ \ \ \ ; \ \ \ \ \ 
Tr_{zz}= Tr^{g}_{zz}+ 4 k^-\left( k_\perp^2 + m^2-\frac{q_x^2}{4} \right)
\ .
\end{eqnarray}

The terms corresponding to the pair production
by the photon in the Breit-frame, are constructed from 
\begin{eqnarray}
B_v(q^2)&=&\imath  \int\frac{d^4k}{(2\pi)^4}
\Lambda(k,P')\Lambda(k,P) 
 \frac{k^-\left( k_\perp^2 + m^2-\frac{q_x^2}{4} \right)}
{(k^2 - m^2+\imath \epsilon)
((k-P)^2 - m^2+\imath\epsilon) 
((k-P')^2 - m^2+\imath \epsilon)}
 \ . \label{bvec}
\end{eqnarray}
 The Cauchy integration in $k^-$, has two pieces
$B^I_v(q^2)$ and $B^{II}_v(q^2)$. In  part (I), the residue is evaluated
at the spectator on-mass-shell pole,  which corresponds
to the integration region of $0<k^+<P^+$. In part (II), 
where $P^+<k^+<P^{'+}$, the pair
term  survives  the limit of $P^{'+}\rightarrow P^+$.
It is obtained from  the pair term
 given in Eq.(\ref{crt}), as
\begin{eqnarray}
B_v^{II}(q^2)= \frac{C^2}{P^+} \int \frac{d^2k_\perp}{(2\pi)^3}
\left( k_\perp^2 + m^2-\frac{q_x^2}{4} \right)
 \sum_{i=2}^{5}\frac{ \ln(f_{i})}
{\prod_{j=2,i\neq j}^{5}(-f_i + f_j)}
\ ;
\label{bii}
\end{eqnarray}
where,
$
f_1  =   k^{2}_\perp+m^2  \ ; \
f_2  =  (P-k)^{2}_\perp+m^2 \ ;  \
f_3  =  (P'-k)^{2}_\perp+m^2 \ ; \ 
f_4 = (P-k)^{2}_\perp+m_{R}^2 \ ; \
f_5  = (P'-k)^{2}_\perp+m_{R}^2 \ .
$

The  matrix elements of $j^+$ are given by
the sum of two terms, one comes from the contribution of
the light-front wave-function
for $0<k^+<P^+$ in Eq.(\ref{jvec}) and the other comes from the pair
term obtained with Eq.(\ref{bii})
\begin{eqnarray}
j^+_{xx}= j^{+I}_{xx}-\frac{q_x^2}{m^2_v} B^{II}(q^2)  ; \
j^+_{zx}= j^{+I}_{zx}-2 \frac{q_x}{m_v}  B^{II}(q^2)  ; \
j^+_{yy}= j^{+I}_{yy} ; \
j^+_{zz}= j^{+I}_{zz}+4  B^{II}(q^2) 
\ .
\label{currv}
\end{eqnarray}

Observe that each matrix element of the current acquires
contribution from the pair term unless $j^+_{yy}$. The  angular
condition in the Cartesian spin basis is,
$\ \ \Delta(q^2)=j^+_{yy}-j^+_{zz}=j^{+I}_{yy}-j^{+I}_{zz}-
4  B_v^{II}(q^2) \ = 0  $  \cite{to93}.
It is zero
since the matrix elements of the current $j^+_{ij}$ of Eq.(\ref{jvec}) 
have the correct transformation properties for rotation around
the $x$-direction. Note that the 
pair terms not only affect the angular
condition but each of the matrix elements of the current.

The violation of the angular condition is a consequence 
of taking into account
 only the contribution of the residue of the pole of the spectator particle
in the Cauchy integration of the $k^-$ momentum, as has also been shown in a
recent numerical study in this model \cite{pach97}.
The violation of the angular condition is  given by,
$\ \ \Delta^I(q^2)=j^{+I}_{yy}-j^{+I}_{zz}=
4  B^{II}(q^2)$ \ . 
The  rotational symmetry of the "good" component of the 
current for spin one particle is valid if the
 pair creation process that survives in the Breit-frame
is included in the computation of the matrix elements. 
The good component of the current receives contribution 
from the pair term for $q^+=0$. 

In summary, we have calculated the pair production terms
in the electromagnetic current of a composite particle,
as a result of the integration of $k^-$ in the  loop-momentum 
of the triangle diagram for the photon absorption process, in the limit of
$q^+$ vanishing.
The $j^-$ component of the electromagnetic current of a composite boson
in the light-front
coordinates is calculated. The residue
associated with the  virtual pair creation process
by the photon in $j^-$, survives in the above limit and it 
is responsible for keeping the covariance properties of the current. 
In one example, 
we also derive the pair terms which are important for maintaining
the rotational symmetry of $j^+$, for a composite spin one
particle,  showing that it is not possible to leave out such contribution
without violating the angular condition in the Breit frame.

\section*{Acknowledgments}
We thank to Prof. S.J. Brodsky for helpful discussions.
This work was supported in part by the Brazilian agencies 
CNPq, CAPES and FAPESP. It also was supported by 
 Deutscher Akademischer Austauschdienst  
and Funda\c c\~ao Coordena\c c\~ao de Aperfei\c coamento
de Pessoal de N\'\i vel Superior (Probral/CAPES/DAAD project 015/95).

\end{document}